\documentclass[%
 reprint,
 superscriptaddress,
 amsmath,amssymb,
 prl,
floatfix,
]{revtex4-2}
\usepackage{amsmath}
\usepackage{graphicx}
\usepackage{dcolumn}
\usepackage{bm}

\usepackage{tikz}
\usepackage{color}
\usepackage{braket}

\begin{document}

\title{Nearly quantized Born Effective charges as probes for the topological phase transition in the Haldane and Kane-Mele models}

\author{Paolo Fachin}
\affiliation{Dipartimento di Fisica, Università di Roma La Sapienza, Roma, Italy}%
\author{Francesco Macheda}
\affiliation{Dipartimento di Fisica, Università di Roma La Sapienza, Roma, Italy}
\author{Paolo Barone}%
\affiliation{CNR-SPIN, Area della Ricerca di Tor Vergata, Via del Fosso del Cavaliere 100,
I-00133 Rome, Italy}%
\affiliation{Dipartimento di Fisica, Università di Roma La Sapienza, Roma, Italy}%
\author{Francesco Mauri}
\affiliation{Dipartimento di Fisica, Università di Roma La Sapienza, Roma, Italy}%

\newcommand{\citeSupp}[0]{Note1}

\begin{abstract}
We propose a new approach to study the transition between different topological states, based on the assessment of the vibrational resonances in infrared spectra. We consider the Haldane and Kane-Mele models finding that Born effective charges are nearly quantized, with a discontinuous jump concomitant with the topological phase transition. In particular, Born effective charges display a finite value in the trivial phase and a null one in the nontrivial one. This is rooted in the connection between Born effective charges and electronic Berry curvature at the band edges.
Finally, at the topological phase transition of the Haldane model, we also
observe a nearly quantized jump of the chiral splitting of the
zone-center phonon frequencies, induced by time-reversal symmetry breaking.
\end{abstract}

\maketitle

\textit{Introduction---} Chiral phonons have recently been observed in non-equilibrium phases of topologically trivial states by using circularly polarized light carrying angular momentum \cite{doi:10.1126/science.adi9601,Ueda2023,doi:10.1126/science.aar2711, PhysRevLett.127.186403}.
An equilibrium, intrinsic, chiral nature of phonons originates by the breaking of time-reversal symmetry \cite{PhysRevB.108.134307}. Here, the molecular Berry curvature induces an anomalous contribution to the phonon eigenvalue equation acting as an effective classical Lorentz force, that breaks the energetic equivalence between differently polarized phonons at zone center \cite{PhysRevB.105.064303, PhysRevLett.130.086701, PhysRevLett.126.225703, PhysRevX.14.011041}. Intrinsic chiral phonons have been related to the nontrivial topology of the Haldane model \cite{PhysRevB.105.064303}, a prototype for the Quantum Anomalous Hall (QAH) insulating state \cite{QAH_review}. Indeed, chiral phonon splittings have been proposed as experimental markers for transitions to topological phases not protected by time-reversal symmetry
\cite{doi:10.1126/sciadv.adj4074}. On the other hand, the detection of topological nontrivial states in systems displaying time-reversal, such as the Quantum
Spin Hall (QSH) states \cite{PhysRevLett.95.226801, PhysRevLett.96.106802}, has been based so far on either charge transport measurements probing the longitudinal resistance in Hall bar geometry \cite{doi:10.1126/science.1133734, doi:10.1126/science.1148047,doi:10.1126/science.aan6003} or discriminating conductive edge states from insulating bulk by directly imaging the local conductivity \cite{ Tang_wte2_natphys2017,jia_wte2_prb2017, reis_bismuthene_science2017, shi_wte2_sciadv2019, Kandrai2020, shumiya_natmat2022, bampoulis_germanene_prl2023}. The challenge of experimental identification of different topological states \cite{Weber_2024} motivated theoretical proposals for alternative detection methods, e.g., measuring collective excitations such as plasmons \cite{PhysRevLett.112.076804, PhysRevLett.119.266804} or the Ruderman-Kittel-Kasuya-Yosida interaction with magnetic impurities \cite{Duan2018}. More recently discontinuous changes of piezoelectric response in 2D time-reversal invariant systems \cite{Yu2020,PhysRevResearch.4.L032006} as well as the Hall viscosity in 2D Chern insulators \cite{PhysRevB.92.165131} have been theoretically suggested as direct probes of topological transitions. 

In this work, we propose a different measurable quantity to detect the transition between trivial and nontrivial topological phases of matter, namely the Born effective charges. Quantifying the polarization change due to atomic displacements, Born effective charges can be experimentally accessed, e.g., from the mode oscillator strengths of vibrational resonances in infrared spectra \cite{born1988dynamical,RevModPhys.73.515, PhysRevLett.68.3603,PhysRevB.68.220510,PhysRevLett.103.116804,PhysRev.157.429, Marchese2024,PhysRevB.58.6224}.
We first investigate their properties for the Haldane model where time-reversal is broken by complex next-nearest-neighbor hoppings \cite{PhysRevLett.61.2015}. We find that Born effective charges display nearly quantized values in the trivial phase, while they almost vanish in the nontrivial one, at odds with chiral phonon splitting displaying an opposite dependence on topology \cite{PhysRevB.105.064303}. This behaviour is well understood via a low-energy expansion of the Haldane model, where the Born effective charge is determined by the difference of Chern numbers evaluated at non-equivalent $\pmb{\textrm{K}}$ and $\pmb{\textrm{K'}}$ points, while the chiral phonon splitting depends on their sum. Superimposing two Haldane models with opposite next-nearest-neighbor hopping restores time-reversal symmetry, thus preventing the occurrence of chiral phonon splitting in the whole phase diagram; still, Born effective charges display quantized jump between different phases. A realistic realization of a topological system retaining time-reversal symmetry is the Kane-Mele model \cite{PhysRevLett.95.226801} that includes a Rashba coupling term induced either by a perpendicular electric field or by the interaction with a substrate, for which we numerically confirm that the Born effective charge can be used as a marker to individuate the onset of the $\mathcal{Z}_2$ topological order, corresponding to the QSH insulator \cite{PhysRevLett.95.146802}. Among the recently proposed materials realizing the Kane-Mele model, germanene has been shown to host a QSH state at experimentally accessible temperatures, with a topological transition to a trivial state induced by a critical perpendicular electric field $E_{z,c}\sim 1.95$ V/nm \cite{bampoulis_germanene_prl2023}. Jacutingaite, a natural occurring layered and exfoliable mineral, has also been predicted to host a large-gap QSH state when in  monolayer form, also tunable by an applied perpendicular field \cite{PhysRevLett.120.117701}. Our proposal could as well be relevant for the experimental investigation of 3D weak topological insulators comprising weakly coupled QSH monolayers effectively described by the Kane-Mele model, whose topological surfaces are usually not cleavable and hard to access using standard surface-sensitive techniques \cite{HasanKane_review,pauly_natphys2015}. 
\\
\textit{Berry curvatures---}  The relationship between topological properties, Born effective charges and chiral phonons can be rationalized in terms of Berry curvatures, encoding the geometric properties of electronic wavefunctions in a crystal. In a periodic crystal the electronic and nuclear degrees of freedom can be decoupled following the Born-Oppenheimer approximation. The nuclei are located at their positions $\pmb{u}_{s}(\pmb{R})=\pmb{R}+\boldsymbol{\tau}_{s}$, where $\pmb{R}$ is a Bravais lattice vector and $\boldsymbol{\tau}_{s}$ indicates the position of the atom $s$ within the cell. The electrons are then described in a mean field framework 
by a single-particle Hamiltonian $H_{\pmb{k}}$, $\pmb{k}$ being the quasi-momentum, which parametrically depends on the atomic positions.
For a generic couple of parameters $(\zeta,\lambda)$, we define the Berry curvature of the occupied manifold as \cite{Vanderbilt_2018, Grosso_2013, RevModPhys.82.1959, RevModPhys.66.899}
\begin{align}
        &\Omega_{\zeta\lambda }(\pmb{k})=-2\textrm{Im}\sum_{n}^{\textrm{occ}}{\braket{{\partial_\zeta \upsilon_{n\pmb{k}}}|{\partial_\lambda \upsilon_{n\pmb{k}}}}}\nonumber \\
    &=-2\textrm{Im}\sum_{n}^{\textrm{occ}}\sum_{m}^{\textrm{empty}}\frac{\bra{\upsilon_{n\pmb{k}}}\partial_\zeta H_{\pmb{k}}\ket{\upsilon_{m\pmb{k}}}\bra{\upsilon_{m\pmb{k}}}\partial_{\lambda}H_{\pmb{k}}\ket{\upsilon_{n\pmb{k}}}}{(\varepsilon_{n\pmb{k}}-\varepsilon_{m\pmb{k}})^2},
    \label{eq:Omzl}
\end{align}
where $\varepsilon_{n\pmb{k}}$ and $\ket{\upsilon_{n\pmb{k}}}$ are the eigenvalues and eigenfunctions of $H_{\pmb{k}}$, and the sum over $n$ and $m$ run on the manifold of occupied and empty bands, respectively. If $\zeta$ and $\lambda$ are atomic positions, then Eq. \ref{eq:Omzl} determines the molecular Berry curvature of the system. As shown in Ref. \cite{PhysRevB.105.064303} for the Haldane model, the molecular Berry manifests as a non-local effective magnetic field in the equations of motion of the ion \cite{cc082398dfa9424b87ec93ba7b5c4305}, inducing a chiral splitting of the zone-center phonon frequencies. The anomalous contribution to the phonon eigenvalue equation at zone center is, in the limit where the phonon frequency is much smaller than the band gap \cite{PhysRevLett.126.225703}, proportional to
\begin{align}
       F_{s\alpha,r \beta}&=\frac{\hbar^2}{\sqrt{M_sM_r}} \frac{A}{(2\pi)^2} \int_{\textrm{BZ}} d^2\pmb{k}  \Omega_{u_{s\alpha}u_{r \beta}}(\pmb{k}),
       \label{eq:Fcomp}
\end{align}
where $r,s$ are atomic indexes and $\alpha, \beta=x,y$ are Cartesian coordinates, $M_{s/r}$ the ion masses and $A$ is the unit cell area. $u_{s\alpha}$ is an atomic displacement equal in each cell of the crystal, so that the $\pmb{R}$ dependence is dropped.

Beside the atomic positions, $H_{\pmb{k}}$ is also parametrically dependent on the quasi-momentum. If we identify $\zeta$ and $\lambda$ with the quasi-momentum, Eq. \ref{eq:Omzl} then describes the electronic Berry curvature. This is related to nontrivial topological properties, such as the QAH Conductivity, via the total Chern number \cite{Vanderbilt_2018, RevModPhys.82.1539}
\begin{align}
    C=\frac{1}{2\pi} \int_{\textrm{BZ}} d^2\pmb{k}\Omega_{k_xk_y}({\pmb{k}}).
    \label{eq:Ccomp}
\end{align}
\\
The Born effective charges can be decomposed as $Z^B_{s,\alpha\beta}=Z_{\textrm{ion}}\delta_{\alpha\beta}+Z^*_{s,\alpha\beta}$ where $Z_{\textrm{ion}}$ is the contribution due to the rigid ionic charge displacement. In \textit{ab initio} framework the latter term coincide with the charge of the ion \cite{Bistoni_2019}, while in a tight-binding approach it contains both the ionic charge and the static electronic density contribution, as discussed in the Supplemental Material \cite{supplementary}.  
The mixed derivative where $\zeta$ is the quasi-momentum and $\lambda$ is the atomic position is instead related to the $Z^*_{s, \beta \alpha}$ in the same limit where the phonon frequency is much smaller than the band gap \cite{PhysRevB.47.1651}
\begin{align}
     Z^*_{s, \beta \alpha}=\frac{A}{(2\pi)^2} \int_{\textrm{BZ}} d^2\pmb{k}  \Omega_{k_\alpha u_{s \beta}}(\pmb{k}).
     \label{eq:crossder}
\end{align}
As shown in the Supplemental Material \cite{supplementary}, $Z_{\textrm{ion}}$ is small with respect to $Z^*_{s, \beta \alpha}$ in the trivial phase and it does not depend on the electronic topological state. Therefore, we neglected this contribution focusing only on the $Z^*_{s, \beta \alpha}$, which are referred in the following as the Born effective charges. Great attention has been devoted to the study of the molecular and electronic Berry curvatures for materials displaying topological properties. In this context, the mixed derivative of Eq. (4) has been analyzed mostly in connection with the piezoelectric response mediated by the electron-strain coupling \cite{Yu2020,PhysRevResearch.4.L032006}, while little attention has been given to Born effective charges of topological states, that will be the focus of this work. 
\\
\textit{Haldane Model---} We consider the Haldane model, described by a tight-binding spinless Hamiltonian on a 2D honeycomb lattice with two atoms per unit cell \cite{supplementary}, one atomic orbital per atom and one electron in each orbital 
\begin{align}
       H_{\textrm{el}}&=\frac{\Delta}{2}\sum_{i }l_i c_{i}^\dag c_{i} +t_1 \sum_{\langle ij \rangle}c_{i}^\dag c_{j}+it_2\sum_{\langle \langle ij \rangle \rangle}l_{ij} c_{i}^\dag c_{j}.
\end{align}
$c_{i},c_{i}^\dag$ are the electronic creation and annihilation operators, where $i$ is a short-hand notation indicating both the cell and the sublattice index. $\Delta/2$ is the on-site energy, $t_1$ and $t_2$ are real quantities and $l_i={1,-1}$ is the sublattice index of the site $i$. $\langle\rangle$ means sum on first nearest neighbours and $\langle\langle\rangle\rangle$ on the second ones. Each term in the sum over $\langle\langle\rangle\rangle$ enters with a sign $l_{ij} = 1,-1$, determined accordingly to the direction of the vector connecting the sites \cite{Vanderbilt_2018,PhysRevLett.61.2015}. We adopt the lattice parameter and  $t_1$ hopping parameters of graphene, i.e. $a=2.46$~\AA~ and $t_1=3.4$eV \cite{PhysRevB.84.035433}, and the atomic masses are the ones of carbon, for which we set $M_s=M$.

\begin{figure}
    \centering
    \includegraphics[width=0.5\textwidth]{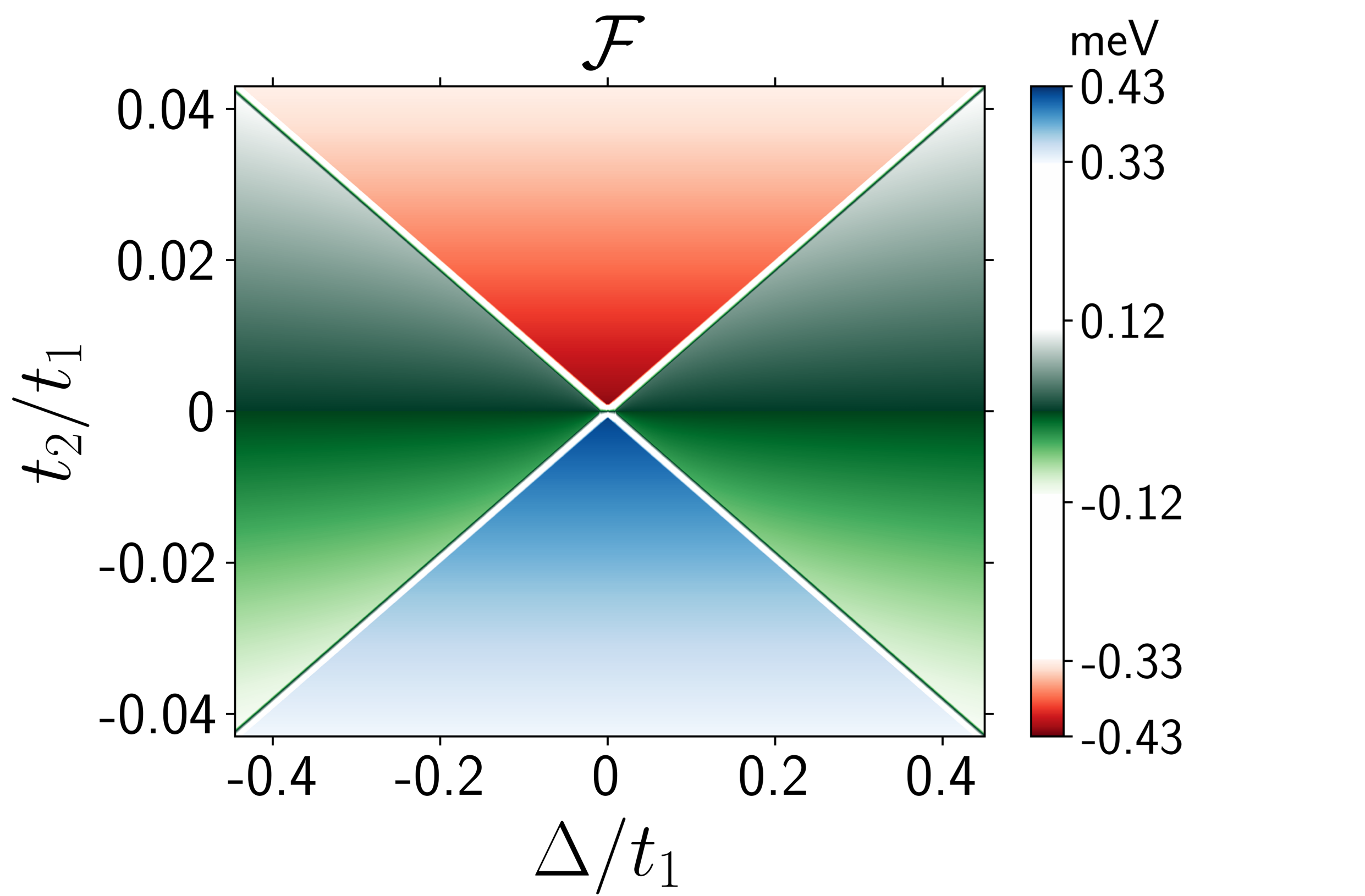}
    \includegraphics[width=0.52\textwidth]{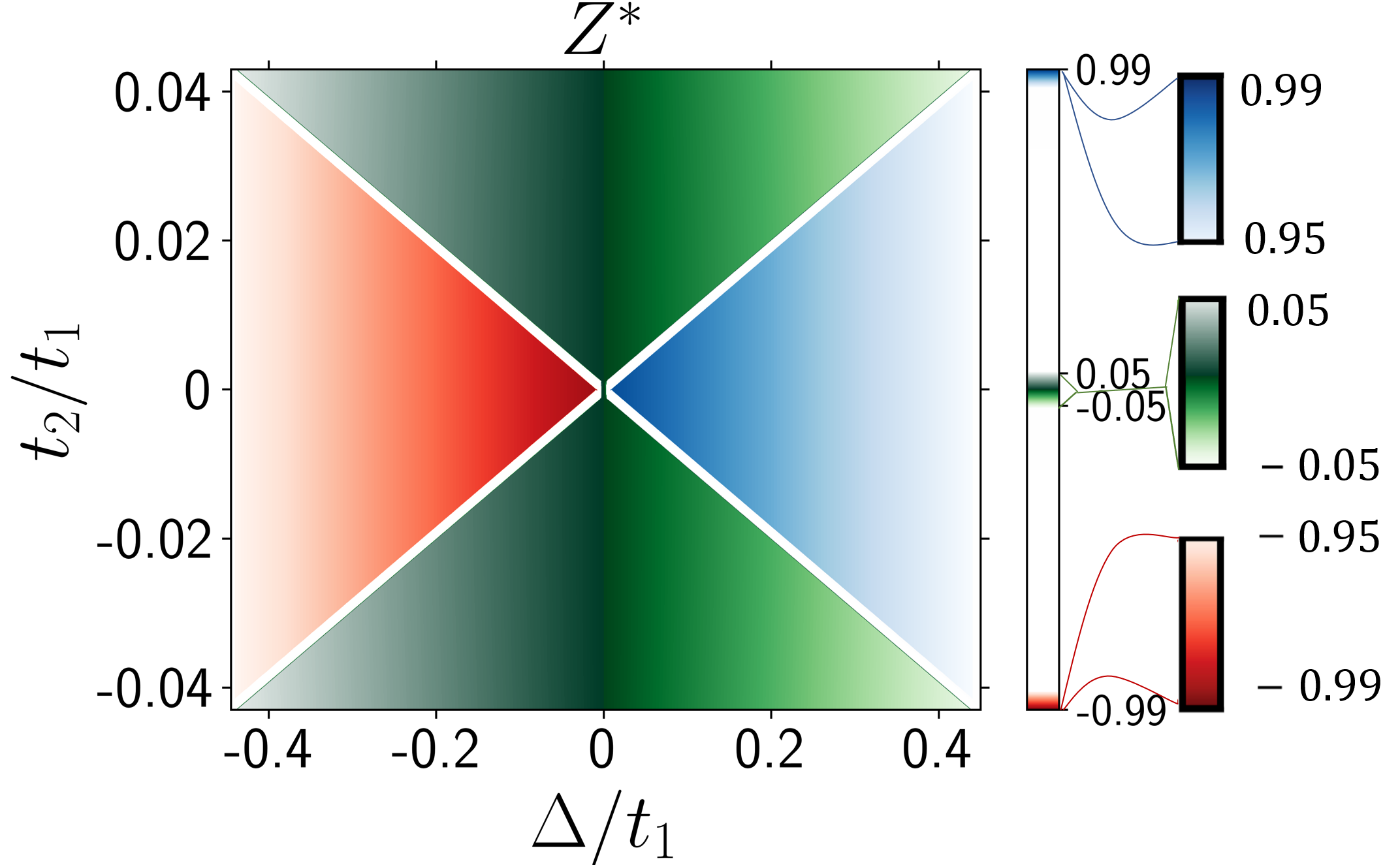}
    \includegraphics[width=0.5\textwidth]{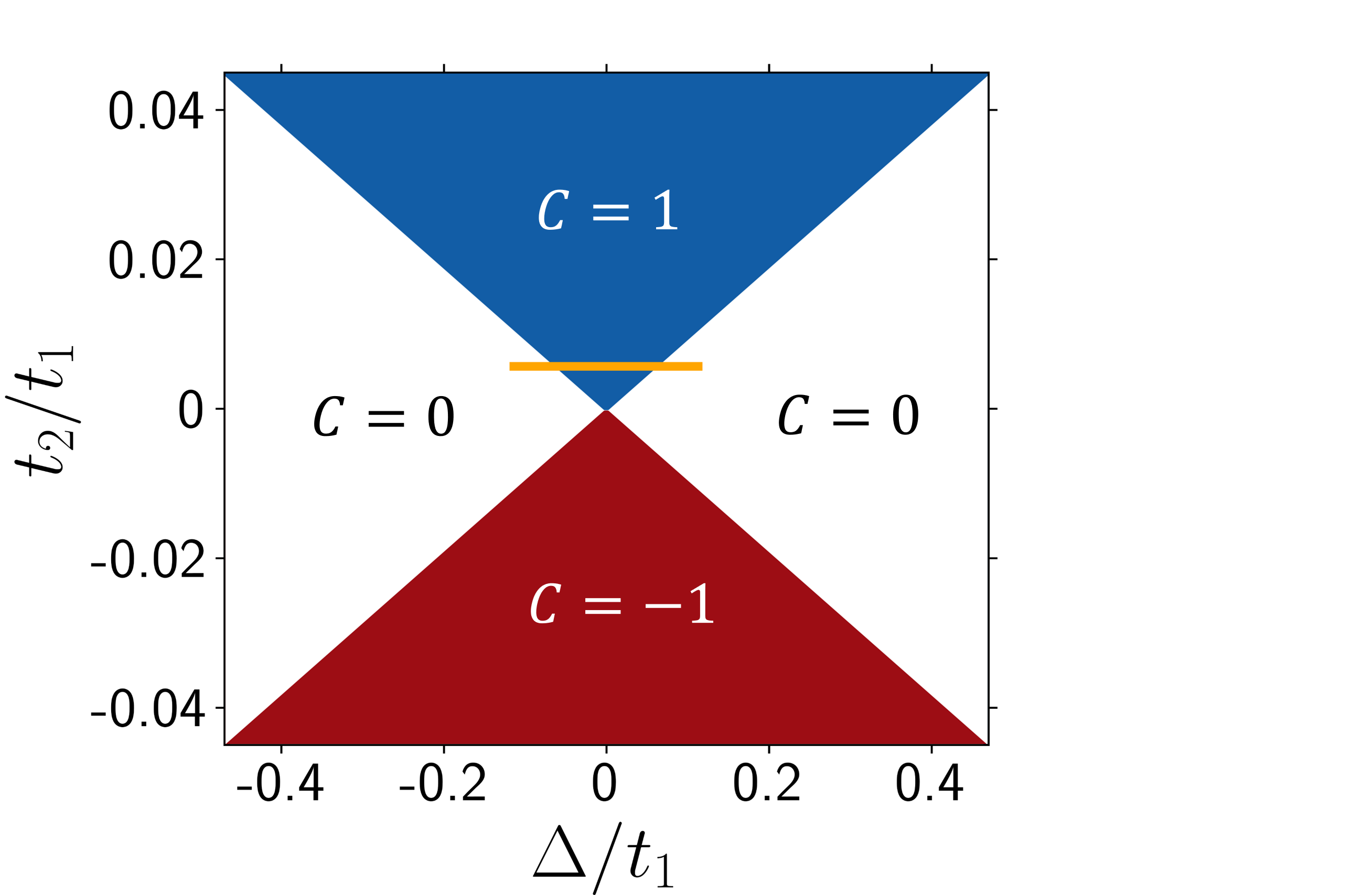}

   \caption{ (Upper) $\mathcal{F}$ and (middle) $Z^*$ display nearly quantized values in topologically different zones of the phase diagram of the Haldane model in the plane ($ \Delta/t_1, t_2/t_1$) for $t_1=3.4$ eV. (Lower) Analytical topological phase diagram of the Haldane model.  The horizontal orange line corresponds to the region of parameter space further expanded in Fig. \ref{fig:KM}.
   }
    \label{fig:Berry_BEC_mappe_Haldane}
\end{figure}

The model describes a two-band system with broken time-reversal symmetry due to the imaginary next-nearest-neighbors hopping term $it_2$. One band is occupied and one is empty, so that we can easily drop the summation over electronic states. The topological phase diagram of the Haldane model in the parameter space ($\Delta/t_1,t_2/t_1$) is represented in the bottom panel of Figure \ref{fig:Berry_BEC_mappe_Haldane}, where the trivial and nontrivial phases are identified by the total Chern number being $C=0$ and $C=\pm 1$, respectively, as computed via Eq. \ref{eq:Ccomp}. The phases are separated by a metallic state, where the gap closes. The coupling between electrons and lattice vibrations is obtained considering the variation of the hopping constants following an atomic displacement. We assume that only the $t_1$ term contributes to the electron-phonon coupling, and that the variation of $t_1$ depends only on longitudinal bond stretchings. The electron-phonon interaction for zone-center lattice vibrations can then be quantified by the coupling parameter $\xi$ \footnote{In the notation of Ref. \cite{Bistoni_2019}, the coupling constant $\xi$ is $\beta^{e-l}/b_0^2$, $b_0$ being the carbon-carbon distance. }
as \cite{PhysRevB.65.235412,PhysRevLett.93.185503,PhysRevB.90.125414,PhysRevB.84.035433}
\begin{align}
        \delta H_{\textrm{e-ph}}&=-t_1\xi\sum_{\langle ij \rangle}\pmb{b}_{ij}\cdot \left(\delta\pmb{u}_{j}-\delta\pmb{u}_{i}\right) c_{i}^\dag c_{j}.
        \label{eq:eph}
\end{align}
The complete Hamiltonian is $H=H_{\textrm{el}}+\delta H_{\textrm{e-ph}}$, so the derivative of Hamiltonian with respect to the atomic positions correspond to the derivative of the term in Eq. \eqref{eq:eph}.
Here $\pmb{b}_{ij}=\pmb{u}_j-\pmb{u}_i$ is the vector connecting the sites $i$ and $j$, while the coupling constant is set to the value obtained for graphene, $\xi=1.2$ \AA$^{-2}$.
\\
Beside the electronic properties, the nontrivial topology also affects the ion dynamics. In particular, in the nontrivial phases the double degenerate optical phonon modes at the $\boldsymbol{\Gamma}$ point split into two circularly polarized modes with opposite chirality and different energies \cite{PhysRevB.105.064303}. For anomalous contributions that are small with respect to the dynamical matrix, the phonon splitting is given by $\Delta \omega=2|F_{sx,sy}|$ (\cite{PhysRevB.105.064303}). The full  $F_{s\alpha,r\beta}$ matrix further contains the information of which chiral phonon is highest in energy. For this reason, 
we consider for each atom $s$ the quantity
 $F_{sx,sy}=l_s\mathcal{F}$.
As shown in the upper panel of Fig. \ref{fig:Berry_BEC_mappe_Haldane}, $\mathcal{F}$ is almost null in the trivial topological phase, while it displays a finite and fairly constant value in different nontrivial ones, with an almost quantized jump in correspondence with the metallic phase. The sign of $\mathcal{F}$ changes when passing from the $C=1$ to the $C=-1$ phases, i.e., when changing the $t_2$ sign.
Finally, we study Eq. \ref{eq:crossder} for the Haldane model. Due to symmetry and charge neutrality, the Born effective charge tensor can be written as
$Z^*_{s,\alpha\beta}=Z^* l_s \delta_{\alpha\beta}$.
As shown in the central panel of Fig. \ref{fig:Berry_BEC_mappe_Haldane}, $Z^*$ appears as almost quantized, with a nearly vanishing values in the nontrivial phase and a finite value in the trivial one. The value of $Z^*$ changes when reversing the $\Delta$ sign. Interestingly, the nearly quantization of $Z^*$ extends to larger portions of the topological space as compared to $\mathcal{F}$.

The above results are rationalized thanks to a low-energy expansion of the tight-binding model. This approximation holds well for small gap values such that
the Berry curvatures are mostly localized around $\pmb{\textrm{K}}$ and $\pmb{\textrm{K'}}$ points in reciprocal space \cite{Bistoni_2019, supplementary}.
Therefore, the observables are expected to closely follow the predictions of the low-energy model, which we obtain by expanding the Hamiltonian for small quasi-momenta $\pmb{p}$ around the non-equivalent $\pmb{\textrm{K}}$ and $\pmb{\textrm{K'}}$. We refer to the sets of points near $\pmb{\textrm{K}}$ and $\pmb{\textrm{K'}}$ as belonging to different `valleys', labeling them as $D_{\eta}$ with $\eta=\pm 1$. In this description, the coupling between electrons and phonons can be expressed as a function of the `gauge' field
\begin{align}
\pmb{\mathcal{A}}=\sum_{s}\pmb{\mathcal{A}}_s, \quad \pmb{\mathcal{A}}_s=l_s \hat{\pmb{z}} \times \delta\pmb{u}_s.
    \label{eq:gauge}
\end{align}
 $\pmb{\mathcal{A}}$ enters the low-energy Hamiltonian of each valley $\eta$ mimicking a minimal coupling \cite{Bistoni_2019,Bernevig_2013}, as 

\begin{align}
    H_\eta=\hbar v_{\textrm{F}} \left(\eta p_x+\xi \mathcal{A}_{x},p_y+\eta \xi \mathcal{A}_{y},\frac{\Delta/2-\eta 3\sqrt{3}t_2}{\hbar v_{\textrm{F}}}\right) \cdot \pmb{\sigma}^{\textrm{P}},
    \label{eq:low_energy}
\end{align}

where  $\hbar v_{\textrm{F}}=\frac{\sqrt{3}t_1 a}{2}$ and $\pmb{\sigma}^{\textrm{P}}=(\sigma^{\textrm{P}}_x, \sigma^{\textrm{P}}_y,\sigma^{\textrm{P}}_z)$ is the vector of Pauli matrices in the pseudospin space, the two pseudo-spinor components corresponding 
to the amplitude of the  periodic part of the Bloch state $\ket{\upsilon_{n\pmb{k}}}$ on the two atom sites in the unit cell \cite{Grosso_2013}. 

\begin{figure}
    \centering
        \includegraphics[width=0.49\textwidth]{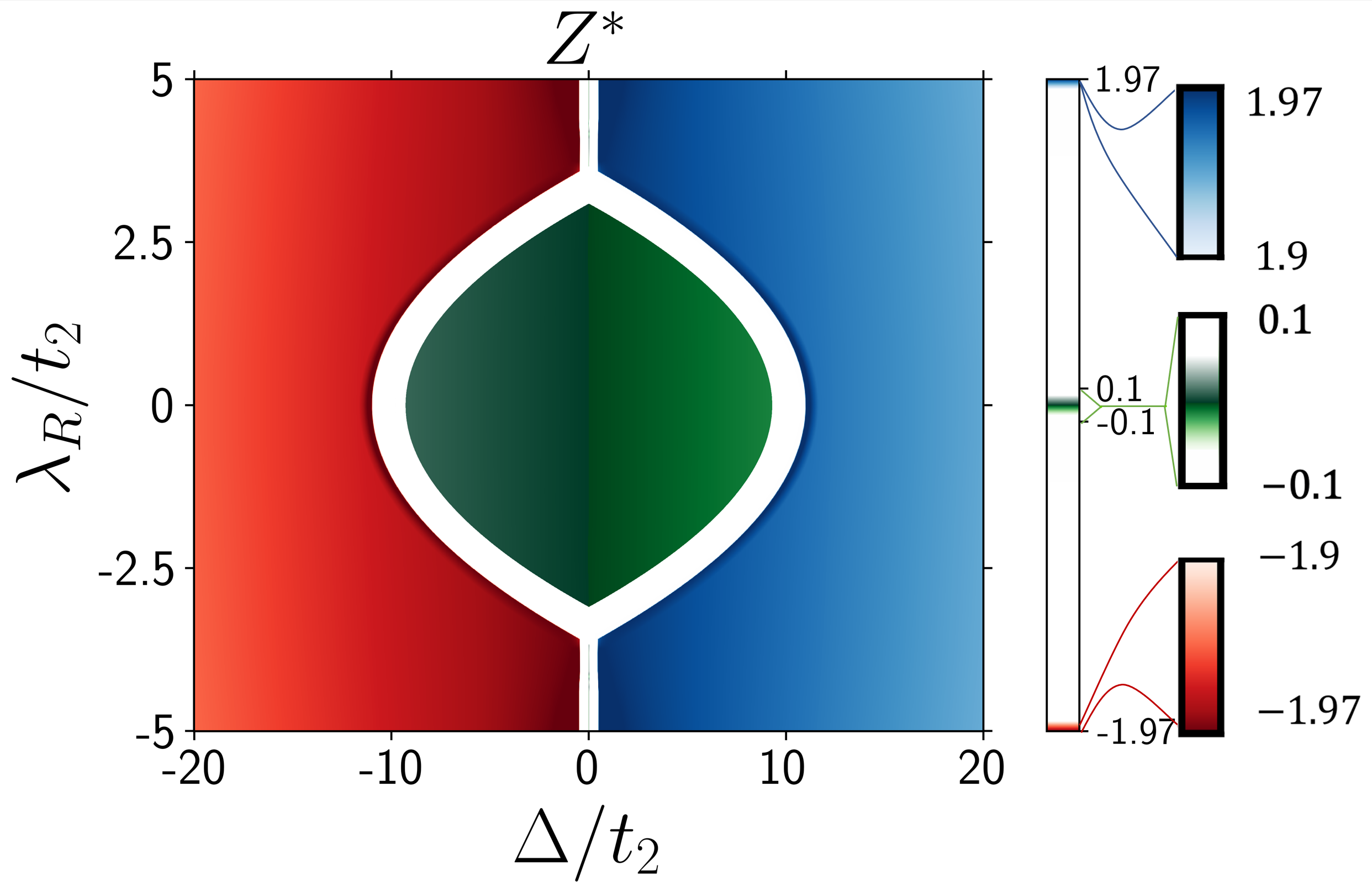}
                    \includegraphics[width=0.49\textwidth]{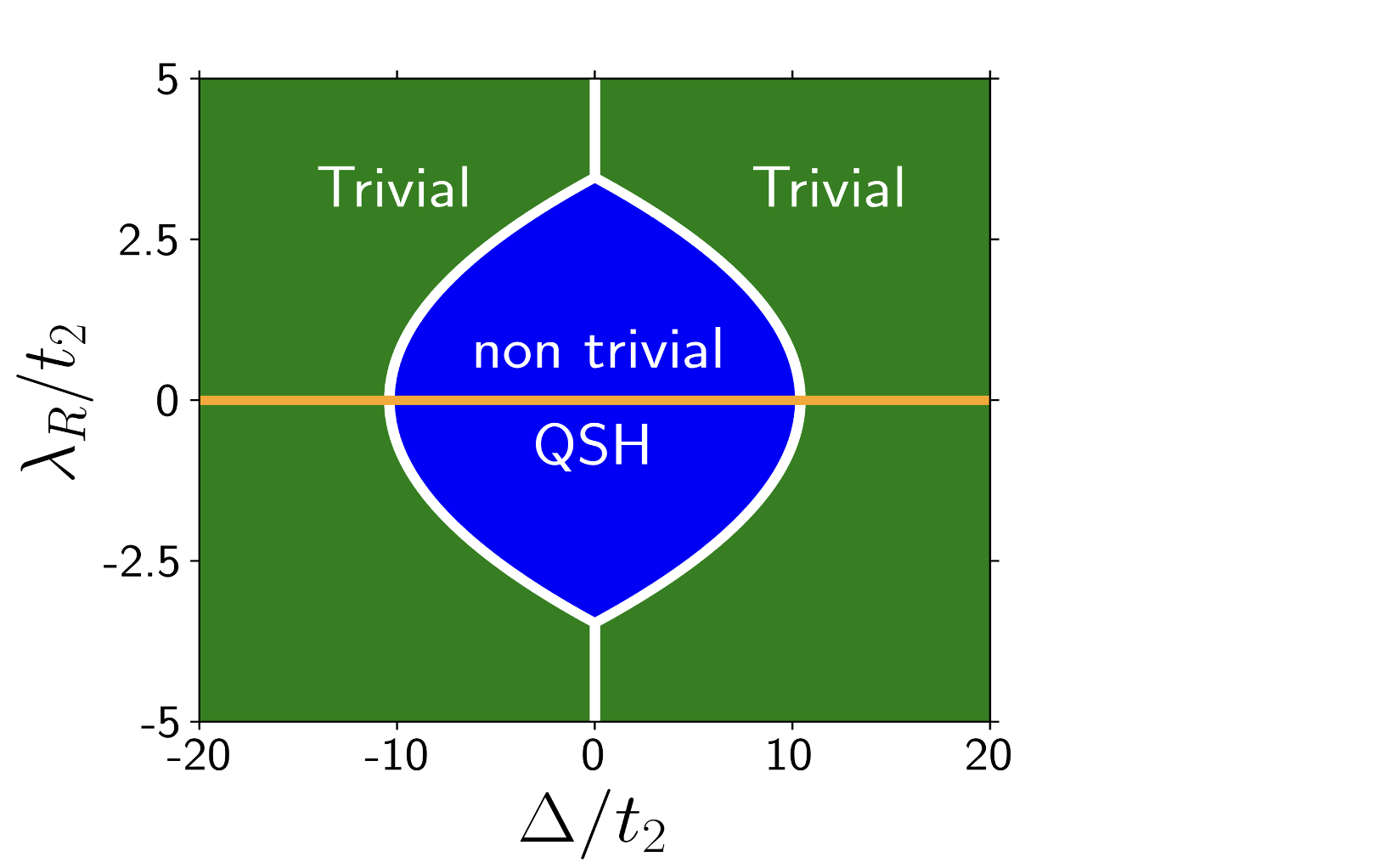}
    \caption{(Upper) $Z^*$ with nearly quantized values in the topologically different regions of the phase diagram of the Kane-Mele model in the  ($\Delta/t_2,\lambda_R/t_2$) plane for $t_2=0.02$ eV. We set $t_2$ as the energy unit to ease the comparison with Ref. \cite{PhysRevLett.95.146802}. (Lower) Numerical topological phase diagram of the Kane-Mele model. The blue region indicates where QSH is realized. The white region numerically identifies band gaps close to metallization. The parameters indicated by the horizontal orange line are the same of the orange line of Fig. \ref{fig:Berry_BEC_mappe_Haldane}, where we adopted $t_1$ as the energy unit; the Rashba coupling is zero along this line,
    and the Kane-Mele model reduces to two time-reversal related copies of the Haldane model.}
    \label{fig:KM}
\end{figure}

For the low-energy model, the integration over all the Brillouin zone is replaced by the sum of the two valleys integrals as $\int_{\textrm{BZ}} d^2\pmb{k} \to \sum_{\eta=\pm1}\int_{D_\eta} d^2 \pmb{p}$. Then, the Chern number of Eq. \ref{eq:Ccomp} is expressed as the sum of two valley Chern numbers $C_\eta=\frac{1}{2\pi}\int_{D_\eta} d^2 \pmb{p}  \,\Omega_{p_xp_y}(\pmb{p})$ as
\begin{align}
   C= \sum_{\eta=\pm1} (\eta)^0 C_{\eta}.
\end{align}
In the trivial phase the valley fraction Chern numbers are $C_\eta=\textrm{sgn}(\Delta)\frac{\eta}{2}$ and cancel out in the sum, while in the topological phase they are both equal to $C_\eta=\textrm{sgn}(t_2)\frac{1}{2}$. To evaluate also $Z^*$ and $\mathcal{F}$, we notice that in the low-energy model of Eq. \ref{eq:low_energy} the derivative with respect to the phonon perturbation transforms in the derivative with respect to the crystalline quasi-momentum as
\begin{align}
\frac{\partial H_{\pmb{p}}}{\partial u_{sx}}=l_s\frac{\partial H_{\pmb{p}}}{\partial A_{sy}}=l_s \eta  \xi {\frac{\partial H_{\pmb{p}}}{\partial p_y}},\\
\frac{\partial H_{\pmb{p}}}{\partial u_{sy}}=-l_s\frac{\partial H_{\pmb{p}}}{\partial A_{sx}}= -l_s \eta  \xi {\frac{\partial H_{\pmb{p}}}{\partial p_x}}.
\end{align}
As a consequence, $Z^*$ and $\mathcal{F}$ are directly related to the electronic Berry curvatures, and thus to the Chern number, as
\begin{align}
Z^*= \frac{A}{2\pi} \xi \sum_{\eta=\pm1} (\eta)^1 C_{\eta}= \textrm{sgn}(\Delta)\frac{A}{2\pi }\xi (1-|C|),\\
\mathcal{F} =-\frac{\hbar^2}{M}\frac{A}{2\pi }\left(\xi\right)^2\sum_{\eta=\pm1} (\eta)^2 C_{\eta}=-\frac{\hbar^2}{M}\frac{A}{2\pi }\left(\xi\right)^2 C.
\end{align}
Moreover, the Born Effective charges provide a direct measure of the valley Chern number $C_V = \sum_\eta \eta\, C_\eta$ \cite{Bistoni_2019}. 
Hence, the nearly quantized properties for the Haldane model are almost all due to the topological properties of its low-energy model, and as such they are quite robust to modification of the electronic band structure, as shown also in the SI \cite{supplementary}. Notice that $C$, $Z^*$ and $\mathcal{F}$ are written in the low-energy model as sums over growing powers of $\eta$. Substituting both phonon derivatives with electronic ones for $\mathcal{F}$ results in an higher degree of approximation in the description of the tight-binding model and, thus, in a less pronounced quasi quantization with respect to the $Z^*$. The result for $\mathcal{F}$ is in agreement with Ref. \cite{PhysRevB.105.064303}. 
\\
\textit{Kane-Mele model---}
In the Haldane model describing spinless electrons the time-reversal symmetry operation on the Hamiltonian corresponds to the change of the sign of $t_2$.  The inversion symmetry, on the other hand, is equivalent to a sign change of the onsite energy $\Delta$. 
As plotted in Figure \ref{fig:Berry_BEC_mappe_Haldane}, both $C$ and $\mathcal{F}$ change sign under the time-reversal symmetry operation, while $Z^*$ is invariant. The behaviour with respect to the inversion symmetry is opposite, since $C$ and $\mathcal{F}$ are invariant while $Z^*$ changes sign.
By combining two Haldane models with opposite sign of $t_2$, i.e., two time-reversed copies of the system, we expect $Z^*$ to sum up and $C$ and $\mathcal{F}$ to vanish. This system actually coincide with the Kane-Mele model without Rashba coupling \cite{PhysRevLett.95.226801,PhysRevB.82.245412}, with a proper identification of the spin-orbit coupling term with $t_2$. Here, the two time-reversed copies of the Haldane model correspond to the two different $\uparrow$ and $\downarrow$ spin components of the system, each copy displaying a Chern number such that $C_\uparrow=-C_\downarrow$ as a consequence of time-reversal symmetry.
In spite of displaying a null total Chern number, the Kane-Mele model 
still presents a topologically nontrivial phase according to the $\mathcal{Z}_2$ classification \cite{PhysRevLett.95.146802}.
It follows that the low-energy expression of the Born effective charge in the Kane-Mele model without Rashba coupling reads:
\begin{align}
    Z^*=\textrm{sgn}(\Delta)\,\frac{A}{2\pi}\xi (2-|C_\uparrow|-|C_\downarrow|).
\end{align}
We then expect $Z^*$ to retain its nearly quantized value discerning between the trivial and the nontrivial $\mathcal{Z}_2$ topological phases, in analogy of what happens for the $\mathcal{Z}$ topological Haldane model.
\\
A full description of the Kane-Mele model requires the inclusion of the Rashba term, mixing spin components. Introducing the coupling constant $\lambda_\textrm{R}$, the Rashba coupling 
reads:
\begin{align}
H_\textrm{R}=&i\lambda_\textrm{R}\sum_{\langle ij \rangle, \rho\rho'}\l_i \, c_{i\rho}^\dag\left[\left(\hat{\pmb{b}}_{ij}\times \hat{\pmb{z}}\right)\cdot\pmb{\sigma}^{\textrm{S}} \right]_{\rho\rho'}c_{j\rho'},
\end{align}
where $\rho,\rho'$ are spin indexes, $\hat{\pmb{b}}_{ij}$ is the nearest neighbour distance normalized to the unity, $\pmb{\sigma}^{\textrm{S}}=(\sigma^{\textrm{S}}_x, \sigma^{\textrm{S}}_y,\sigma^{\textrm{S}}_z)$ is the vector containing Pauli matrices in spin space. Being time-reversal symmetric, the Rashba term preserves the topological properties as long as the system is adiabatically connected to the QSH phase at $\lambda_R=0$, i.e., until the band-gap closes \cite{PhysRevLett.95.146802}. Acting on $\lambda_\textrm{R}$, e.g. by applying a perpendicular electric field, the system can be driven from the topological to the trivial phase and viceversa by crossing the metallic line, as shown in the lower panel of Fig. \ref{fig:KM} for $t_2=0.02$ eV. The corresponding evolution of $Z^*$ is displayed in the upper panel, confirming the expected nearly quantized behavior of the effective charge. We numerically checked the relationship between Born effective charge, valley Chern number and the topological index $\mathcal{Z}_2$, as shown in SI \cite{supplementary}.
\newline
\textit{Discussion and conclusions---} 
We showed that in the Haldane and Kane-Mele models the Born effective charges display an almost quantized jump between different topological phases. Its origin can be ascribed to the electron-phonon coupling entering as a gauge field in the low-energy Hamiltonian that captures the topological phase transition, such that both molecular and mixed Berry curvatures, accounting for chiral phonon splitting and Born effective charges respectively, can be expressed in terms of electronic Berry curvature accounting for the topological character of the band structure. In the considered models Born effective charges can be explicitly related to the valley Chern number, a well defined and meaningful quantity when approaching the transition point as the electronic Berry curvature gets strongly localized around valley points \cite{supplementary}. Nevertheless, a topological transition between two insulating states can be generally described by an effective Hamiltonian of massive Dirac fermions analogous to Eq. (\ref{eq:low_energy}) \cite{PhysRevB.76.205304,PhysRevB.96.125127,Yu2020}, displaying a discontinuous jumps of electronic Berry curvature across the transition point\cite{Bernevig_2013, Vanderbilt_2018}. Under general assumptions it can be shown that the e-ph coupling to optical phonons can always display a gauge-field component when the effective Hamiltonian matrix is defined in a pseudospin space accounting for sublattice degrees of freedom \cite{PhysRevB.100.165427,supplementary}, suggesting that discontinuous jump of Born effective charges could manifest across other topological transitions beyond the models considered here as, e.g., in the (Pb,Sn)Te material class of topologically crystalline insulators \cite{TCI_natcomm2012,TCI2_natcomm2012}. We remark that, at odds with `deformation potential' contributions to electron-phonon interaction, gauge-field terms are unaffected by electronic screening \cite{PhysRevB.90.125414}, hinting to further robustness of the proposed effect. Finally, we emphasize the analogy with the predicted discontinuous changes of piezoelectric response in 2D time-reversal invariant systems, recently proposed as experimental markers of topological transitions \cite{Yu2020, PhysRevResearch.4.L032006}, also relying on well-defined valley Chern numbers and on the electron-strain coupling acting as a gauge field \cite{Bistoni_2019, Droth_prb2016, Rostami2018}.

\textit{Acknowledgments---} We acknowledge the MORE-TEM ERC-SYN project, grant agreement no. 951215. We acknowledge the EuroHPC Joint Undertaking for awarding this project access to the EuroHPC supercomputer LUMI, hosted by CSC (Finland) and the LUMI consortium through a EuroHPC Regular Access call.
\bibliography{biblio}
\nocite{Winkler_book,https://doi.org/10.1002/qua.560210105, PhysRevB.83.235401,PhysRevB.89.115102, PhysRevLett.129.185902, PhysRevB.107.094308, PhysRevB.109.075420} 
\end{document}